\documentclass[prl,aps,preprint,showpacs,floatfix]{revtex4}
\usepackage{graphicx}
\usepackage{epstopdf}
\usepackage{wrapfig}
\usepackage{subfigure}
\usepackage{caption}
\usepackage{amsmath}
\usepackage{ctable}
\usepackage{amsmath}
\usepackage{microtype}
\usepackage[version=3]{mhchem}
\usepackage[dvipdfm]{hyperref} 
\usepackage{datetime}
\begin{document}
\title{Dynamical complexity in the quantum to classical transition} 
\author{
Bibek Pokharel$^{a}$, 
Peter Duggins$^{(a)}$, 
Moses Misplon$^{(a)}$, 
Walter Lynn$^{(a)}$, 
Kevin Hallman$^{(a)}$, 
Dustin Anderson$^{(a)}$, 
Arie Kapulkin$^{(b)}$, 
and 
Arjendu K. Pattanayak$^{(a)}$} 
\date{\today}
\affiliation{(a) Department of Physics and Astronomy, Carleton College,
Northfield, Minnesota 55057
\\
(b) 128 Rockwood Cr, Thornhill, Ont L4J 7W1 Canada
} 
\begin{abstract}
We study the dynamical complexity of an open quantum driven double-well 
oscillator, mapping its dependence on effective Planck's constant
$\hbar_{eff}\equiv\beta$ and coupling to the environment, $\Gamma$. 
We study this using stochastic Schrodinger equations, semiclassical 
equations, and the classical limit equation. We show that (i) the 
dynamical complexity initially increases with effective Hilbert 
space size (as $\beta$ decreases) such that the most quantum systems 
are the least dynamically complex. (ii) If the classical limit is 
chaotic, that is the most dynamically complex (iii) if the classical 
limit is regular, there is always a quantum system more dynamically
complex than the classical system. There are several parameter regimes 
where the quantum system is chaotic even though the classical limit 
is not. While some of the quantum chaotic attractors are of the same
family as the classical limiting attractors, we also find a quantum 
attractor with no classical counterpart. These phenomena occur in
experimentally accessible regimes.
\end{abstract}
\maketitle

Investigations of how quantum behavior limits to (significantly different) 
classical dynamics inform quantum information, control, and technology. 
The limiting behavior is particularly intriguing for nonlinear systems: 
A generic classical nonlinear system (Hamiltonian or dissipative) 
can display dynamically complex behavior. The dynamical complexity 
includes in particular exponentially sensitive dependence on how 
precisely the initial condition is specified, visible in the 
distance $d$ between initially close trajectories growing with 
time $t$ as $d(t) \approx d(t=0) e^{\lambda t}$. Here $\lambda$ 
is the Lyapunov exponent, one aspect of the Kolmogorov complexity of
the dynamics. Closed quantum dynamics are entirely (quasi)-periodic,
with no dynamically complex behavior (no positive $\lambda$), yielding 
the paradox that chaotic classical Hamiltonians are a singular limit. 
Investigating this paradox and the singular limit greatly improved 
our understanding of semiclassical dynamics and also of the 
properties of quantum spectra and eigenfunctions as they relate to 
classical chaos~\cite{chaology}. 

However, nonlinear quantum systems are sensitive to environmental 
effects and their accurate description must include consideration 
of decoherence\cite{decoherence}. As such, further understanding of 
nonlinear effects in quantum dynamics emerged in studies of stochastic 
quantum trajectories, through conditioned measurement records or via 
the Quantum State Diffusion (QSD)\cite{percival} approach. Early
work\cite{spiller,brun,ota} showed that a classical phase-space 
chaotic attractor was recovered smoothly from quantum 
Poincar\'e sections for $\langle \hat Q \rangle$ and 
$\langle \hat P \rangle$ for a sufficiently small scaled Planck's 
constant $\hbar_{\rm eff} (\equiv \beta$ below) and it 
disappeared slowly as $\beta$ increases, 
Positive quantum $\lambda$ have indeed been calculated~\cite{habib}, 
establishing that quantum complex dynamics exists. We are far from 
a complete understanding of this phenomenon however. Among many
conjectures is the belief~\cite{finn} that quantum effects supress 
dynamical complexity in general and chaos and moreover, that 
quantum chaos exists only for systems which are classically chaotic. 
This dynamical complexity of quantum systems is of great interest to 
the field of quantum computing as well\cite{burgarth}.

In this Letter, we investigate the dynamically complexity (
quantified below) of the open (dissipative) quantum Duffing 
system. Starting from high $\beta$ (extreme quantum behavior) 
we find that the quantum system is always more dynamically complex 
than completely regular behavior (as for a classical periodic orbit). 
As $\beta$ decreases so that the system becomes somewhat less quantal or
more classical, the dynamical complexity of the system initially 
increases. This behavior transitions smoothly to the classical limit
as $\beta$ continues to decrease. However, since the classical behavior 
can be either be (I) chaotic attractors of high dynamical complexity or 
(II) periodic orbits (dynamically simple behavior) we get two distinct 
results. For the case (I) the classical behavior is the most dynamically 
complex and for (II) there is a non-monotonic transition in that there 
is always a quantum system with complexity greater than the classical 
limiting behavior. The non-monotonic transitions include `anomalous' 
examples of quantum chaos for systems with a periodic classical limit.  
Further, while some of these anomalous quantum chaotic attractors resemble 
the classically chaotic attractors at neighboring parameters, there are 
examples where the anomalous quantum chaotic attractors are unrelated 
to any classical dynamics.  The last, to the best of our knowledge, is 
the first example of uniquely quantum chaotic attractors. 

To understand this, we start by summarizing the classical behavior and 
its complexity; our particular system evolves as 
\begin{equation}
\ddot{x} + 2\Gamma \dot{x} + \beta^2 x^{3} - x = \frac{g}{\beta} \cos(\Omega t),
\label{Eq:cDuff}
\end{equation}
representing a unit mass in a double-well potential, with 
dissipation $\Gamma$ and a sinusoidal driving of amplitude $g$ and
frequency $\Omega$, and is termed the Duffing oscillator. 
Equation~(\ref{Eq:cDuff}) is invariant under $\beta\to\Lambda\beta; x 
\to \frac{x}{\Lambda}$ for any positive real $\Lambda$. 
We vary the dissipation parameter $\Gamma$ and see the range of behaviors 
shown in the bifurcation diagram Fig.~(1). This is constructed as 
follows: (a) fix $\Gamma$, (b) evolve an arbitrary initial condition 
over 200 periods of the driving, (c) discard the transients 
($10$ drive periods) and record $x$ ($p$ can also be used with no 
change in the analysis) at $t=2n\pi$ for all $n$ and 
(d) plot {\em all} the $x(t=2n\pi)$ results against $\Gamma$. 
These (and all results below) use a driving amplitude of $g=0.3$.
\begin{figure}[h]
\includegraphics[scale=0.18]{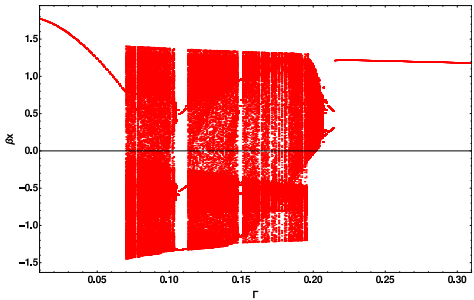}
\caption{Classical bifurcation diagram as a function of $\Gamma$}
\label{fig.bifurcation}
\end{figure}
A chaotic orbit has $x(t)$ wandering densely in phase space and 
looks like a solid line, while a regular (periodic) orbit yields 
just a few points. As $\Gamma$ changes, chaotic 
attractors that persist with $\Gamma$ look like a solid band, and periodic 
orbits that persist yield simple curves, as in Fig.~(2). Also visible 
are periodic-doubling cascades and transitions to (and from) chaos
with $\Gamma$. The embedded regularity within generally chaotic 
dynamics results from higher-order symmetries emerging in the 
evolution equation. Such behavior, including the period doubling 
cascades, is generic in chaotic systems, though not always 
present\cite{sanders}. All our calculations have non-zero $\Gamma$ 
since $\Gamma =0$ is a singular Hamiltonian limit particularly for 
the quantum system.

This is quantified by the dynamical complexity $K =\lambda + \Gamma$ 
plotted against $\Gamma$ in Fig.~(2); we also see $K$ for various 
quantum systems, discussed below. $K$ is the computational complexity
quantifying the precision needed for the initial condition to get an accurate 
trajectory at some asymptotic time. For our system, phase-space lengths 
contract as $\Gamma$ whence $K$ correctly captures that given a globally 
attractive periodic orbit, zero information is needed about the initial 
condition to compute the entire future orbit. $K$ is thus a measure 
of information needed about initial conditions for relative to this 
baseline for periodic orbits in dissipative systems. Finally, any system 
with $K$ above the $\Gamma$ diagonal has positive $\lambda$ and is hence 
formally chaotic. We see that initially as $\Gamma$ increases away from 
zero, the classical dynamics remain simple with $K=0$. 
Around $\Gamma_1\approx 0.070$ there is an abrupt transition to complex 
behavior ($K>0$), which persists until $\Gamma_2\approx 0.210$, except 
for the many small windows of regular behavior most prominently around 
$\Gamma\approx 0.110$ for example. The figure thus shows the presence 
of one family of chaotic (complex) dynamical attractors, with the 
system switching between this and regular (simple) behavior. At 
low damping the global simple attractor is a periodic orbit moving in 
synchrony with the driving across both wells; the driving compensates 
for the dissipation. $\Gamma_1$ marks when 
the dissipation is large enough such that the orbit does not always 
not make it over the central potential barrier. This yields 
the potential for chaos: given two nearby orbits, one may cross 
the barrier while the other does not, leading to drastically different 
evolution. For $\Gamma > \Gamma_2$ the damping is so high that the 
dynamics are restricted entirely to one of the wells, and does 
not yield complex behavior. We see that the $K>0$ `hill' of
complex dynamics arises from dynamics interacting with the central 
barrier of the potential; i.e., this `complexity hill' marks a 
dynamical phase transition between simple behavior of differing
symmetry. Thus, classically simple ($K=0$) dynamics corresponds to: 
(i) At low dissipation $\Gamma < \Gamma_1$, trivial double-well 
periodic orbits, (ii) at high dissipation $\Gamma > \Gamma_2$, 
trivial single-well periodic orbits, and 
(iii) at intermediate dissipation $\Gamma_1  < \Gamma < \Gamma_2$.  
high order periodic orbits in $\Gamma$ windows of various sizes. 
\begin{figure}[htbp]
  \centering
  \includegraphics[width=.8\linewidth]{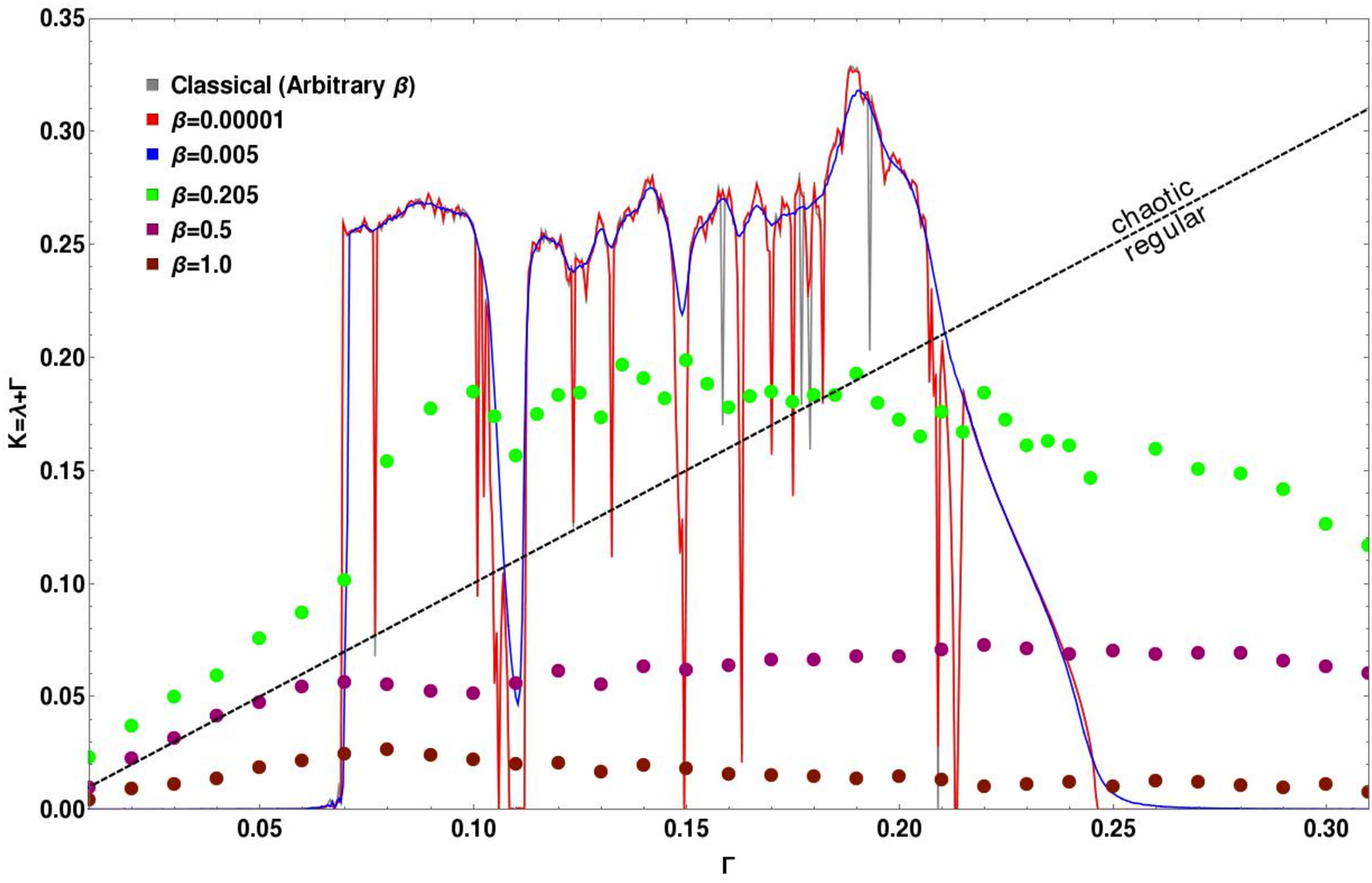}
\caption{ Complexity $K = \lambda_q + \Gamma$ vs $\Gamma$
for various $\beta$.  }
\label{fig.LyapGamma}
\end{figure}

The quantum dynamics are described by the Ito equation\cite{percival} for 
$|\psi\rangle$ as
\begin{eqnarray}
|d\psi\rangle &=& -\frac{i}{\hbar}\hat H |\psi\rangle dt
+ \sum_j (\hat L_j - \langle \hat L_j \rangle )|\psi \rangle
d\xi_j\nonumber\\
&+& \sum_j \bigg (
\langle \hat L_j^\dagger \rangle \hat L_j
-\frac{1}{2}\hat L_j^\dagger \hat L_j
-\frac{1}{2}\langle \hat L_j^\dagger \rangle \langle \hat L_j\rangle
\bigg )
|\psi\rangle dt
\label{sse}
\end{eqnarray}
\noindent where $\hat H$ is the Hamiltonian and the $\hat L_j$ are
the Lindblad operators representing the action of a zero-temperature
Markovian environment\cite{percival}. The $d\xi_j$ are independent 
normalized complex differential random variables satisfying 
$M(d\xi_j) =0; M(d\xi_j d\xi_{j'}) =0; M
(d\xi_jd\xi^*_{j'})=\delta_{jj'}dt$ where $M$ represents the mean over
realizations. 
This can be made dimensionless\cite{brun,kapulkin,ota} with the Hamiltonian 
$\hat{H_\beta}=\hat{H}_{D}+\hat{H}_{R}+\hat{H}_{ex}$ 
and a Lindblad operator $\hat{L}$ for Eq.~(\ref{sse}), 
where $\hat{H}_{D} = \frac{1}{2} \hat{P}^{2}+\frac{\beta^{2}}{4}\hat{Q}^{4}
-\frac{1}{2} \hat{Q}^{2},
\hat{H}_{R} = \frac{\Gamma}{2} \left(\hat{Q} \hat{P} + \hat{P}
\hat{Q} \right),
\hat{H}_{ex} = - \frac{g}{\beta} \hat{Q} \cos(\Omega t),
\hat{L} = \sqrt{\Gamma} \left( \hat{Q} + i\hat{P} \right).$ 
Both the quantum nonlinearity and the dissipation scale with $\Gamma$,
via $\hat{L}$. $\beta$ is dimensionless and related to the original 
system's parameters of mass $m$, length $l$, and natural frequency 
$\omega_0$ as $\beta^{2} = \frac{\hbar}{ml^{2}\omega_{0}}$ and serves as 
$\hbar_{\rm eff}$ determining the scale (and degree of `quantumness')
of the system\cite{brun, ota}. That is, increasing $\beta$ amounts to
a smaller system needing fewer quantum states to describe the dynamics, 
and $\beta\to 0$ is thence the classical limit. 
Recently analysis\cite{liqi} shows that experimentally accessible
nano-electromechanical systems (NEMS) are well described by this model 
with parameters of recent experiments within range of the phenomena we report.

To fully map the quantum-classical transition, we also consider the 
semiclassical dynamics (\cite{halliwell,ota} ) 
derived from the evolution equations for 
$\langle\hat Q\rangle, \langle\hat P\rangle$. 
These include their dependence on second moment terms 
$\sigma_{QQ},\sigma_{PP},\sigma_{PQ}$ as well as the environmental 
term $d\xi$, where 
$\sigma_{AB} = \langle (\hat A^\dagger - \langle\hat A\rangle^*)
(\hat B - \langle\hat B\rangle) \rangle$. The second moments 
are themselves time dependent and depend on higher order terms 
like $\sigma_{QQQ}$ etc. An infinite hierarchy\cite{andrews} 
of time dependent moments corresponds to the full quantum dynamics. 
Various approximations exist\cite{schieve}; we truncate at the 
second order, leaving us 
\begin{align}
dx &=& p dt + 2 \sqrt{\Gamma} ( (\mu - \frac{1}{2}) d \xi_R - R d \xi_I)
\\
dp &=& (- \beta^2 (x^3 + 3 \mu x) + x - 2 \Gamma p + \frac{g}{\beta} cos
(\omega t)) dt\nonumber\\
 &+& 2 \sqrt{\Gamma} ( R d \xi_R - ( \kappa -\frac{1}{2}) d
\xi_I)
\end{align}
\begin{align} 
\frac{d \mu}{d t} & = 2 R + 2 \Gamma (\mu - \mu^2 - R^2 + \frac{1}{4}
)\\
\frac{d \kappa}{d t} & = 2 R(- 3 \beta^2 x^2 + 1)+ 2 \Gamma ( - \kappa -
\kappa^2 - R^2 + \frac{1}{4}) \\
\frac{d R}{d t} & =\mu (- 3 \beta^2 x^2 +1 ) + \kappa - 2 \Gamma R (\mu
+ \kappa),
\end{align}
where $\langle \hat Q\rangle \equiv x, \langle \hat P \rangle \equiv p
\equiv \dot{x}$, 
$R \equiv \frac{1}{2} (\sigma_{QP} + \sigma_{PQ})$,
$\sigma_{QQ} \equiv \mu$ and $\sigma_{PP} \equiv \kappa$.
Equations~(4,5) contain the classical Eq.~(\ref{Eq:cDuff}), 
along with quantum terms. The first of these depends on $\mu$ 
and arises from $\hat H$ being nonlinear. Other terms emerge 
via $\hat{L}$ and depend on $\sqrt{\Gamma}d\xi$ and 
$\sigma (=\mu,\kappa,R)$. As $\beta \to 0$ the range of $x,p$ 
increases as $\frac{1}{\beta}$ while $\sigma$ and $d\xi$ terms remain 
unchanged in scale (of order unity). Thus, as $\beta$ decreases, 
the product terms like $\simeq\sigma d\xi$ quickly become relatively 
negligible, followed by the separate $\sigma,d\xi$ terms, and these 
equations reduce to the classical Eq.~(\ref{Eq:cDuff}) for a point particle, 
identified with $(x,p)$, the quantum centroid. 

This scaling of various moments is consistent with and justifies 
the semiclassical truncation. The semiclassics can be variously
validated, most simply by comparing with the full quantum evolution. 
For a given parameter set, as $\beta$ decreases from some value 
(say $\simeq 0.1$ for this system), once the semiclassics agree 
with the full quantum dynamics at some $\beta_{\rm sc}$ they can only 
{\em improve} in accuracy for $\beta < \beta_{\rm sc}$. 
This depends on system parameters in interesting ways, 
but is empirically found to be greater than $\beta\simeq 0.01$. 
The full quantum calculations scale in computational difficulty 
as roughly $\exp(\beta^{-3})$. We have deployed two implementations 
(the QSD library\cite{percival} using a moving basis technique 
efficient at higher $\Gamma$ and low $\beta$, and a fixed basis 
XMDS2\cite{xmds2} method which performs better at lower 
$\Gamma$ and higher $\beta$), along with semiclassics at even 
lower $\beta$, to cover the full range reported. 

To compute the quantum $\lambda$, a fiducial $\psi_{fid}$ and 
perturbed $\psi_{per}$ with $\psi_{per }$ created by a 
random small unitary kick applied to $\psi_{fid}$, are 
evolved with identical noise realizations. Then 
$\lambda = \lim_{t \to\infty} 
\lim_{\Delta x(0), \Delta p(0) \to0} \frac{1}{t}  
log_2 \left(\sqrt{\frac{(\Delta x(t))^2+(\Delta p(t))^2}
{(\Delta x(0))^2+(\Delta p(0))^2}}\right)$
where $\Delta x(t) = 
\langle \hat Q \rangle_{\psi_{fid}} - 
\langle \hat Q \rangle_{\psi_{per}}$ and similiarly $\Delta p(t)$.
As is canonical\cite{wolf}, the limit of infinite time and zero
initial distance is replaced by the perturbed trajectory being 
periodically reset to remain within the linear deviation regime, 
and the logarithmic rate of change of distance before every reset is 
averaged. To ensure convergence, the $\lambda$ are computed over 
$3000$ periods of the driving, and averaged over multiple noise 
realizations.

Figure~(2) summarizes our central results, comparing $K$ for 
(i) the classical system, (ii) $\beta_1=0.00001$, the extreme 
semi-classical limit, (iii) $\beta_2 =0.02$, (iv) $\beta_3 =0.205$, 
(v) $\beta_4=0.5$, and finally (vi) an extreme quantum version at 
$\beta_5=1.0$.  The quantum $K$ versus $\Gamma$ results can be 
summarized as follows: (a) They broadly resemble the classical and 
differ increasingly with increasing $\beta$. (b) For any 
$\Gamma$, $K$ starts low at $\beta=1.0$ and increases with 
decreasing $\beta$ at least through $\beta\approx 0.2$.  
(c) If the classical limit is chaotic, $K$ is the greatest for the
classical system, (e) For all other cases, $K$ is non-monotonic 
with $\beta$ such that there is always a quantum system more 
dynamically complex than the classical limit.

The initial increase in $K$ as $\beta$ decreases from $1.0$ results 
from the increase in size of the effective Hilbert space of the dynamics. 
That is, at $\beta\approx 1.0$ the dynamics are described by two or 
so quantum states, hence with low dynamical complexity and low 
$K$ (though not as simple as classical periodic orbits); 
the number of states increases with $\beta^{-1}$, as numerically 
verified by us. This holds true even at $\Gamma$ values corresponding 
to the `complexity hill', which flattens ($K$ decreases) as 
$\beta$ increases since the central barrier matters less due to 
tunneling effects. 
As with the classical dynamics, the $\Gamma$ dependence at a given 
$\beta$ is quite complicated. Generically (this is most visible at
intermediate $\beta=0.205$), $K$ increases with $\Gamma$
initially due to the $\Gamma$ dependence of the nonlinearity in Eq.~(2). 
We then have the jump in $K$ as with the classical system for the
`complexity hill', followed by a slow decrease as $\Gamma$ increases; 
this last arises from tunneling effects delaying the dissipative
transition to single-well dynamics for quantum systems. 
In looking at the transition of the quantum behavior to the classical 
limit, for small enough $\beta$ the quantum results trace the classical 
curve, including the return to $K=0$ at the low and high $\Gamma$ values,
{\em and} in the narrow windows in the intermediate $\Gamma$ regime. 
The number of $\Gamma$ windows where the classical and quantum results
disagree decrease as $\beta$ decreases; the $\beta =0.00001$ results 
illustrate this, but we emphasize that we have recovered the classical 
limit for {\em all} $\Gamma$ values. Considering in particular chaos 
($\lambda$) rather than $K$, therefore, the quantum-to-classical 
transition can be anomalous in being regular-to-chaotic instead of the 
usual chaotic-to-regular; this happens in all three regimes where 
classically regular behavior exists.

The anomalous behavior in the intermediate dissipation regime 
(e.g. for $\Gamma=0.110$) is understood semiclassically by 
comparing Eqs.~(5-9) with the classical Eq.~(1) when the spread 
variables $(\mu,\kappa,R)$ are relatively small. The dynamics of 
these spread variables when coupled to the centroid motion can 
destroy the delicate classical periodic orbits, with the effects 
increasing with $\beta$. The narrowest windows -- where the 
smallest change to the dynamical equations sends the system back 
into chaos -- are naturally the most difficult to reproduce 
quantum mechanically or semiclassically; the perturbative effect of 
the quantum variables destroys the classical regularity in these windows. 
The quantum (semiclassical) attractor in these situations has 
essentially the same $K$ as the classical chaotic attractor family 
and is visually essentially identical.  This is clear in the 
Poincare sections shown in Fig.~(3), constructed by periodically 
recording $(x(t), p(t))$ once every driving period, i.e. at $t=2n\pi$ 
for all $n$. We see regular classical behavior destroyed by the 
quantum mechanics which `recovers' the classically chaotic attractor 
(to be compared with, for example, the classical attractor in 
Refs.~\cite{brun, kapulkin, ota}). Similiar anomalous chaos also 
obtains for $\Gamma \approx 0.209$. Physically, the classical 
trajectory is regular since it is dissipatively constrained to one 
well. As $\beta$ increases quantum tunneling yields a quantum chaotic 
attractor resembling the classical attractor, (see Fig.~(4)) 
with quantum $K$ values essentially the same as for the classical 
attractors. We note that these results also confirm that even at 
$\Gamma=0.3$, intermediate $\beta$ systems have greater $K$ values 
than the classical system even though they are not chaotic\cite{kapulkin}.
\begin{figure}
  \centering
\begin{subfigure}
\centering
  \includegraphics[scale=0.5]{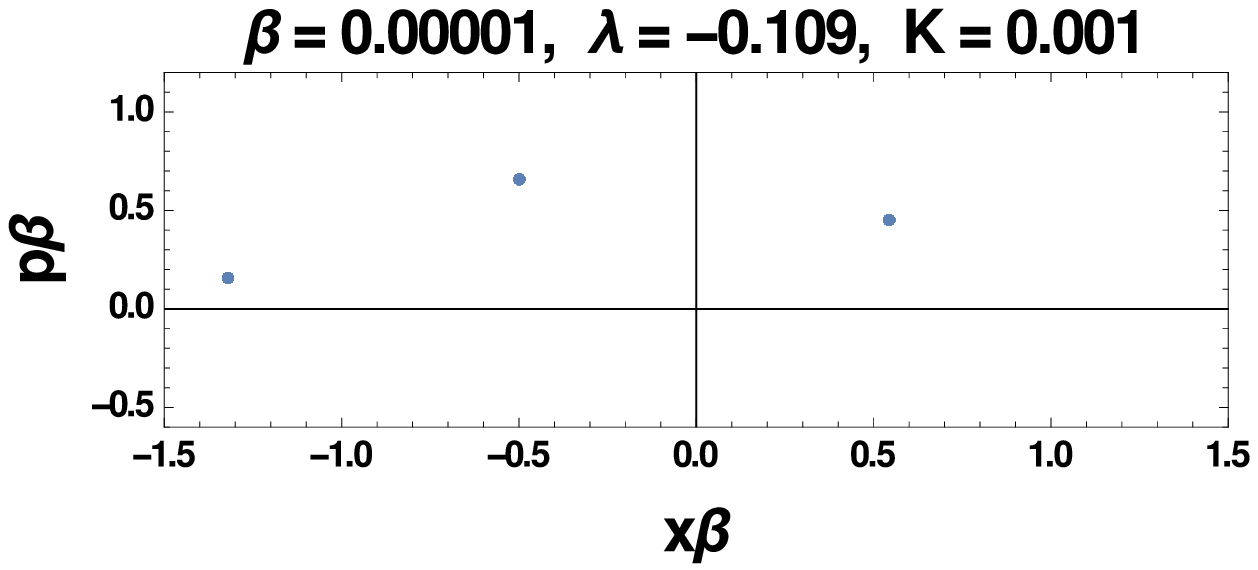}
\end{subfigure}%
\centering
\begin{subfigure}
\centering
 \includegraphics[scale=0.5]{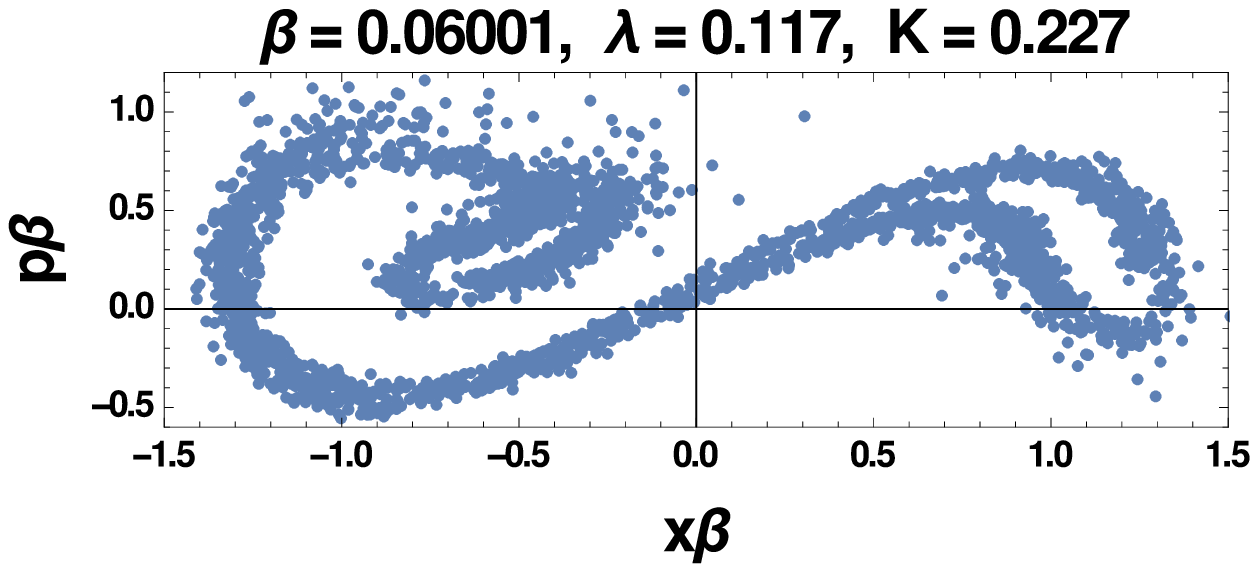}
\end{subfigure}
\caption{Poincare maps for $\Gamma=0.11$}
\label{fig.Gamma110}
\end{figure}

\begin{figure}
  \centering
\begin{subfigure}
\centering
  \includegraphics[scale=0.5]{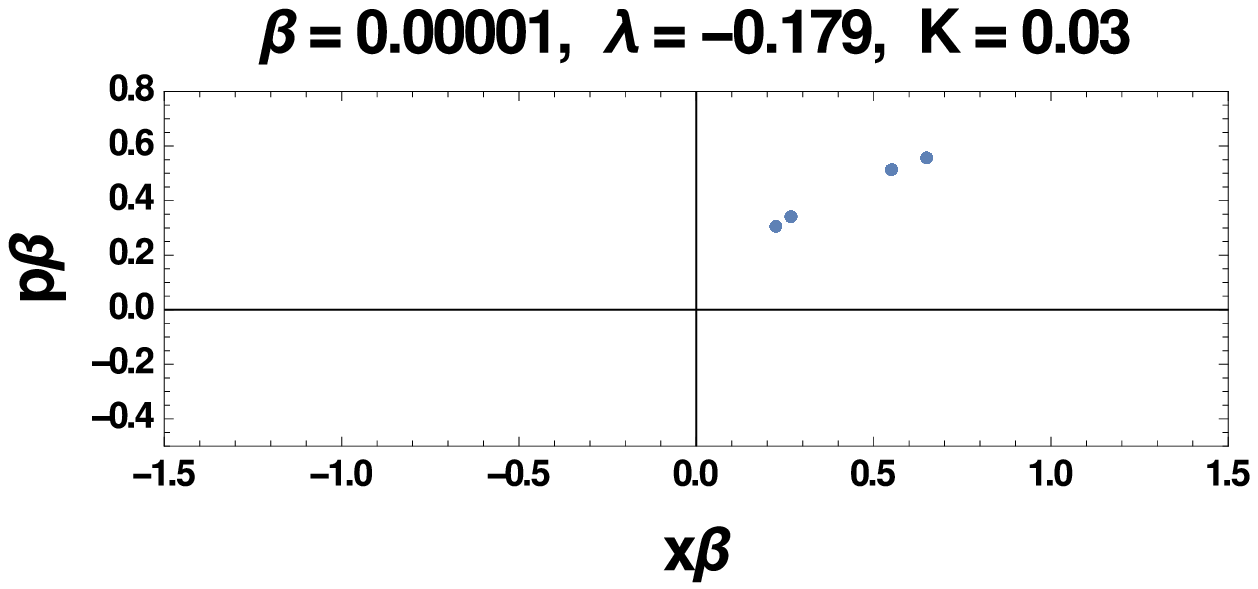}
\end{subfigure}%
\begin{subfigure}
\centering
 \includegraphics[scale=0.5]{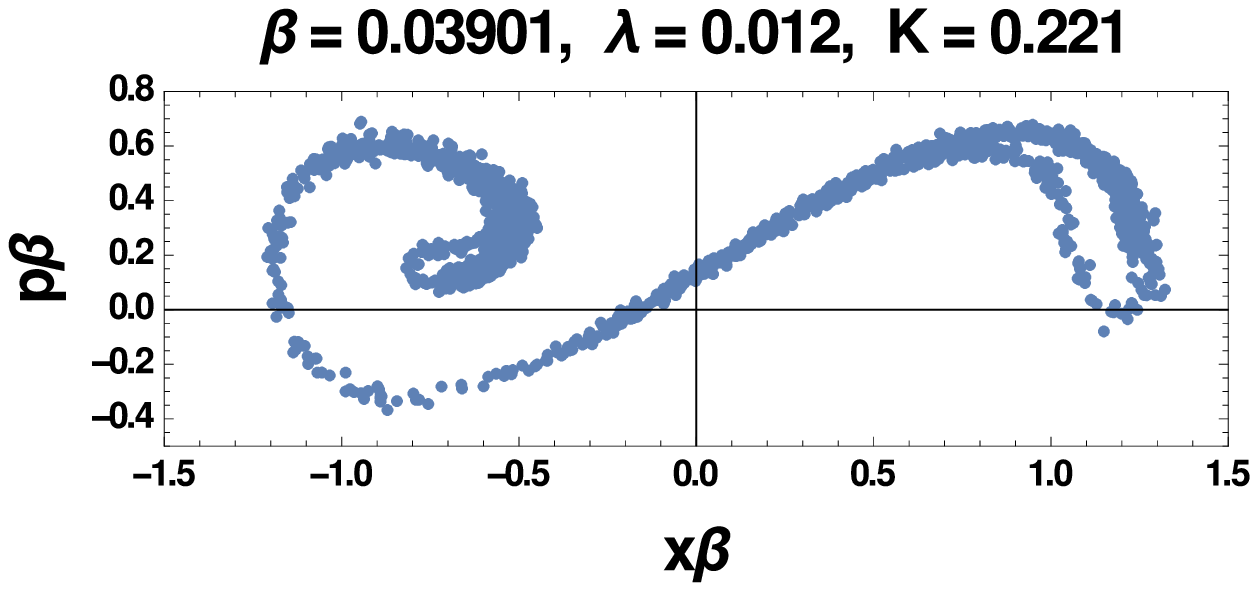}
\end{subfigure}
\caption{Poincare maps for $\Gamma=0.209$}
\label{fig.Gamma209}
\end{figure}

\begin{figure}
  \centering
\begin{subfigure}
\centering
  \includegraphics[scale=0.5]{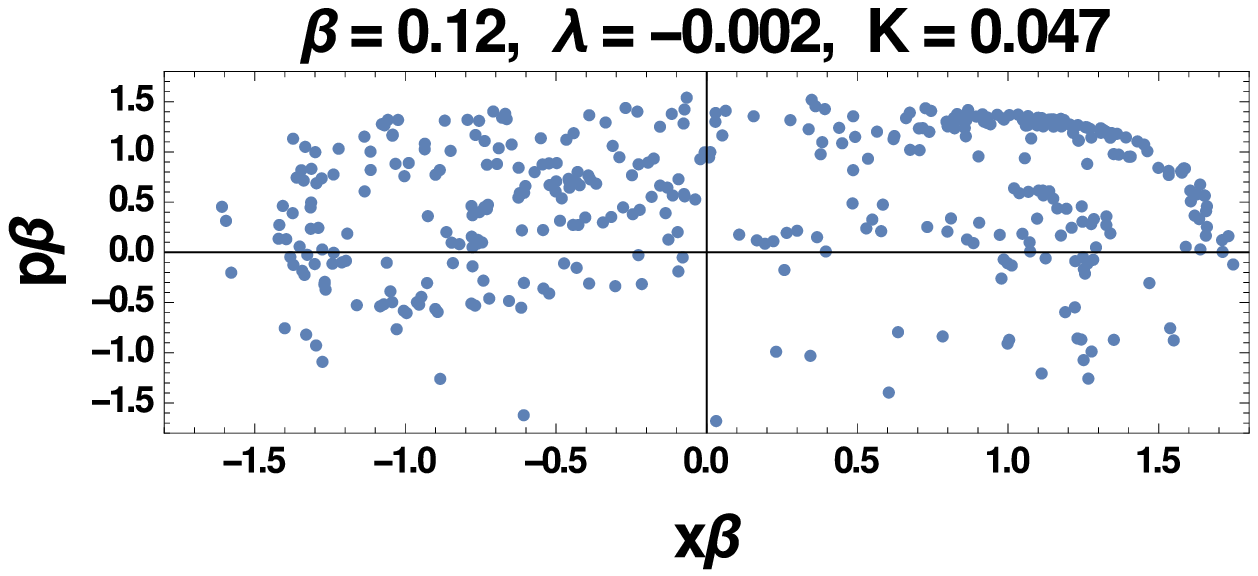}
\end{subfigure}
\begin{subfigure}
\centering
\includegraphics[scale=0.5]{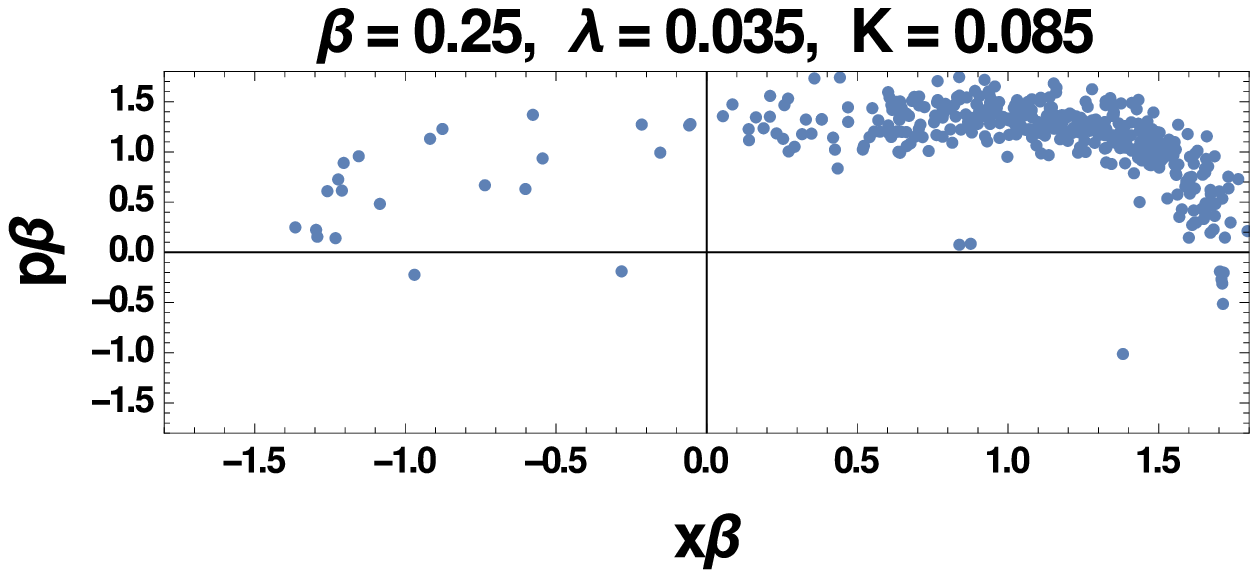}
\end{subfigure}
\caption{Poincare maps for $\Gamma=0.05$}
\label{fig.Gamma050}
\end{figure}

Finally, at low damping $\Gamma < 0.07$ there is also 
quantum chaos (see the $\beta=0.205$ curve) that does not exist 
classically. As evidence that this is not `recovered' classical chaos, 
but is uniquely quantal we note: Both $K,\lambda$ differ substantially 
from that for the other chaotic attractors (classical and quantum),
as does the Poincare section (Fig.~(5)). Also, unlike the other cases, 
the phase-space contracts for the chaotic attractor ($\lambda> 0$) 
compared to when $\lambda <0$. At $\Gamma=0.05$ this transition 
happens as $\beta$ increases beyond $\approx 0.12$. At this 
$\beta$ semiclassics are clearly not valid, but the structure 
of Eqs.~(5-7) for $\mu,\kappa,R$ is instructive. Starting with $\beta$ 
near $0$ and briefly ignoring the $\Gamma$ terms, we have three linear 
equations with a constant of motion ($R^2 -\mu\kappa$). At these
$\Gamma$ values, the $\Gamma$ terms are first order corrections, and 
effectively add more linear terms which cannot engender chaos. However, 
as $\beta$ increases, the $\beta^2 x^2$ term creates a discernible 
nonlinear coupling between Eqs.~(3-4) and Eqs.~(5-7) and breaks the 
integrability. This is essentially the same as for the previous
anomalous chaos, but here it happens at intermediate $\beta$, and does 
not restore a classical-like chaotic attractor. Accurate semiclassics 
in this $\beta$ regime include more terms and equations of higher 
$\beta$ order. However, those higher terms are not necessary for chaos 
since both the semiclassical system Eqs.~(3-7) and the full quantum 
system Eq.~(2) are chaotic in this $\beta$ regime, such that the 
semiclassics are qualitatively valid. 
This is uniquely quantum trajectory chaos, without a corresponding 
classical attractor. The ingredients for this anomaly are sufficient 
quantum nonlinearity (size of Lindblad terms), and intermediate $\beta$ 
(so the Hilbert space is large enough to not be too simple but $\beta$ 
is not so small as to track classical regularity) and this should be 
a generic phenomenon. Interesting avenues for further investigation 
include the dependence of the break point ($\beta_{\rm sc}$) on $\Gamma$ 
and the degree of classical or quantum chaos. We have already seen 
examples where the semiclassics works better for a system with greater 
classical chaos compared to a system that is classically regular, 
contrary to standard intuition.

Our results show that the dynamical complexity across the quantum-to-classical 
transition for open nonlinear quantum systems is substantially richer
than the current intuition that quantum effects cause classically chaotic 
and dynamically complex behavior to transition to quantum regularity and 
simple behavior. The details depend strongly on system particulars. In 
general the quantum evolution is dynamically complex, and this means 
that if the classical limit is regular, the deeper quantum regime may 
display chaos that does not exist in the classical limit. In particular, 
as in the intermediate dissipation regime where the classical system has 
regular behavior corresponding to higher-order dynamical symmetry, the 
classical regular behavior can be difficult to recover -- we have to 
go to larger and larger length scales (smaller $\beta$) to recover 
correspondence. That is, ironically contrary to the standard 
wisdom\cite{chaology}, the correspondence limit can be most difficult 
for certain classically regular systems.  How much of this behavior 
depends on the specific choice of Lindblad operator (that is, the exact 
form of the relevant stochastic Schrodinger equation) is an intriguing 
question currently being investigated\cite{jessica}. It is already 
clear that the broader parameter regimes in question for this and 
other explorations are within reach of current experiments\cite{liqi}. 

A.K.P. is grateful for funds from the HHMI through Carleton and 
the hospitality of J.M. R\"ost at MPIPKS in Dresden, 
G. Mantica at CNCS at the University of Insubria in Como, 
as well as of KITP at Santa Barbara, and helpful conversations with A.
Carvalho.

\end{document}